\documentclass{iopjournal}

\usepackage{amsmath}
\usepackage{amssymb}
\usepackage{xcolor}

\begin{document}

\articletype{Paper} 

\title{Oscillatory Active Brownian Motion:\\
A Minimal Model for Sperm Dynamics}

\author{Adrian Pacheco-Pozo$^{1}$\orcid{0000-0003-2550-4566}, Arturo Matamoros Volante$^{1}$\orcid{0000-0003-1750-3942}, Pilar Ameijeiras$^{1,2}$\orcid{0000-0001-5486-6668}, Mariano G. Buffone$^{2}$\orcid{0000-0002-7163-6482}, and Diego Krapf$^{1,*}$\orcid{0000-0002-2833-5553}}

\affil{$^1$School of Biomedical and Chemical Engineering, Colorado State University, Fort Collins, Colorado 80523, USA}

\affil{$^2$Instituto de Biolog\'ia y Medicina Experimental (IBYME-CONICET), Ciudad Aut\'onoma de Buenos Aires, Argentina}

\affil{$^*$Author to whom any correspondence should be addressed.}

\email{diego.krapf@colostate.edu}


\begin{abstract}
Active biological microswimmers typically combine persistent self-propulsion with cyclic motion generated by flagellar or ciliary beating. However, standard active Brownian motion (ABM) does not explicitly account for these intrinsic oscillations. This limitation is particularly relevant for sperm cells, whose transport depends not only on directional persistence and stochastic reorientation but also on periodic head motion. Here we introduce oscillatory active Brownian motion (OABM), a minimal extension of ABM in which the swimmer orientation undergoes directional diffusion while being modulated by a periodic angular drive. The model yields analytical expressions for experimentally relevant observables, including the time-averaged mean-squared displacement, velocity autocorrelation function, and transverse excursion amplitude. A central prediction is a crossover between two ballistic regimes: short-time motion governed by the swimming velocity and an intermediate regime with a reduced, oscillation-averaged effective velocity determined by the angular beat amplitude. Using trajectories of human sperm, we infer model parameters from standard motility measures. The model quantitatively reproduces both single-cell trajectories and population-level dynamics under control conditions and following induction of hyperactivation. Overall, OABM provides a compact active-matter framework that links measurable flagellar kinematics to coarse-grained transport, enabling a description of sperm motility across different physiological states.
\end{abstract}

\section{Introduction}

Active motion is a defining feature of many biological systems. Self-propelled cells such as motile bacteria, flagellated algae, and spermatozoa transform chemical energy into directed motion, thereby overcoming thermal fluctuations and exploring the environment via persistent trajectories. These cells display a broad range of swimming strategies, from the run-and-tumble dynamics of bacteria \cite{berg1972chemotaxis,Wadhwa2022} to the breaststroke-like motion of biflagellate algae, such as \textit{Chlamydomonas reinhardtii} \cite{Geyer2013,Wei2024}. Despite their biological diversity, they share a common physical signature: motion emerges from the interplay between an underlying active cyclic drive and stochastic fluctuations \cite{Bechinger2016}.
Among microswimmers, sperm cells represent a canonical and biologically important example, since successful navigation to and fertilization of the egg sustain the continuity of life \cite{Gallagher2019,Gaffney2021}. Spermatozoa are highly specialized flagellated cells, whose motility enables them to reach the site of fertilization \cite{Gaffney2011,Vyklicka2020}. Therefore, sperm motility patterns are considered a fingerprint of their fertilizing potential \cite{gomendio2007sperm}. However, sperm are one of the most diverse cell types in nature \cite{hook2020methodological}. For species with internal fertilization, such as mammals, sperm motility is modulated by the conditions encountered in the female reproductive tract, and the motility can change as the cell progresses towards the oocyte \cite{katz1983evolution,buffone2012heads}. Furthermore, this motility is intrinsically heterogeneous: even within a single sample, individual cells exhibit substantial variability in velocity, beat frequency, and directional persistence \cite{Holt2004,Ramon2014}. 

A natural starting point for modeling sperm flagellar propulsion is the framework of active Brownian motion (ABM), which has become a standard coarse-grained model for persistent stochastic motion and has been previously used in the context of sperm cells \cite{Solon2015,Zaferani2023}. In this framework, a swimmer is represented as a self-propelled particle moving at constant velocity along its instantaneous orientation, resulting in a persistent random walk where translational fluctuations and rotational diffusion randomize its trajectory \cite{Romanczuk2012}.  
However, sperm trajectories exhibit a characteristic feature that lies outside the scope of standard ABM: the periodic oscillations generated by flagellar beating \cite{Basu2018}. These oscillations produce a regular side-to-side motion of the sperm head that is superimposed on the overall forward progression of the cell. In contrast, conventional ABM assumes that changes in swimming direction arise solely from stochastic reorientation, leading to smooth persistent trajectories without an intrinsic periodic component. Consequently, ABM cannot distinguish between directional fluctuations that originate from random rotational noise and those that arise from the cyclic motion imposed by the flagellum. This limitation is particularly important because flagellar beating not only shapes the geometry of individual sperm trajectories but can also influence how persistent self-propulsion is translated into long-range transport \cite{Ishimoto2017}.

The need for an extended model becomes particularly evident when studying different physiological states. As mentioned above, the swimming pattern of a sperm cell can change during its journey. A hallmark transformation in mammalian sperm behavior is the transition to hyperactivated motility \cite{ho2001hyperactivation,freitas2017signaling}. The transition into this state is required to allow sperm to penetrate the layers surrounding an egg. In hyperactivated sperm, the lateral displacement of the head increases and the beat frequency changes \cite{Gaffney2011,suarez2003hyperactivated}. These changes arise in part from alterations in the oscillatory component of motion. Therefore, a model that neglects periodic flagellar forces misses an essential component of the swimming dynamics. In hyperactivated sperm, the oscillatory pattern leads to strong lateral forces that play a crucial role in fertilization \cite{suarez2008control,yanagimachi1970movement}. In the female reproductive tract, hyperactivation occurs in response to stimuli present in specific regions of the tract, whereas during in vitro fertilization, it can be induced by properly designed media. 

In this article, we address the sperm motility limitations of ABM by introducing oscillatory active Brownian motion (OABM), a minimal extension of ABM in which the swimmer orientation is subjected to a periodic angular modulation while retaining the conventional rotational-diffusion dynamics. This choice is motivated by the observation that the dominant effect of flagellar beating at the level of sperm-head trajectories is an approximately periodic angular oscillation superimposed on persistent forward motion. Introducing a single oscillation amplitude and frequency constitutes the smallest modification of ABM capable of reproducing experimentally observed head oscillations, oscillatory velocity correlations, and beat-frequency signatures in the power spectrum, while preserving the analytical tractability and physical interpretation of the ABM framework. The resulting model provides a direct bridge between flagellar-beat characteristics and coarse-grained transport properties.
We combine experiments, analytical derivations, and numerical simulations to characterize and validate the OABM model for sperm motility. We measure trajectories of human sperm under control conditions and after treatment with a cyclic AMP analog, 8-Br-cAMP, that induces hyperactivation \cite{Ritagliati2026mot}, extracting standard motility and oscillatory observables such as curvilinear velocity, transverse head motion, beat frequency, mean-squared displacement, and velocity autocorrelation function. We derive analytical expressions for key OABM observables. Finally, we use experimentally inferred parameters to perform numerical OABM simulations at both the single-cell and population levels, comparing control and hyperactivated sperm dynamics. This combined experimental, analytical, and computational approach provides a statistical framework for linking measurable oscillatory kinematics to coarse-grained sperm transport.


\section{The model}
The motion of a sperm cell emerges from the interaction between a rapidly beating flagellum and the surrounding fluid. However, the microscopic details of the flagellar waveform are not typically resolved in measurements of sperm trajectories. Instead, the trajectory of the sperm head consists of a persistent forward progression superimposed with an approximately periodic side-to-side motion. Thus, our goal is to develop a minimal coarse-grained model that captures the dominant kinematic features of the observed sperm trajectories. By focusing on the trajectory level, the model provides a compact description of persistent propulsion, stochastic reorientation, and periodic directional oscillations, without attempting to resolve the detailed mechanics of the flagellum or its hydrodynamic interactions with the surrounding fluid.

We represent the sperm head as an active particle moving with persistent self-propulsion while undergoing two forms of directional change. The first is a  
slowly varying orientation that accounts for fluctuations in the swimming direction and is modeled using the standard ABM framework \cite{Basu2018}. The second is
a periodic modulation arising from the cyclic flagellar beating \cite{woolley2010flagellar}. 
At the trajectory level, this oscillation appears as a cyclic modulation of the swimming direction and can be introduced as
\begin{equation}
\psi(t) = A \cos(\Omega t + \phi).
\label{eq:psi_def}
\end{equation}
where $A$ is the angular amplitude, $\Omega$ the oscillation frequency, and $\phi$ the phase at time zero.
The periodic angular variable $\psi(t)$ should not be interpreted as a mechanical description of the flagellum. Instead, it is a coarse-grained representation of the cyclic modulation of the swimming direction produced by flagellar beating and reflected in the observed sperm trajectory.

We start from the standard two-dimensional ABM \cite{Basu2018}. A swimmer at position $\mathbf{r}(t)=[x(t),y(t)]$ moves at constant velocity $V$ along its instantaneous orientation $\theta(t)$,
\begin{equation}
\begin{aligned}
\dot{x}(t) &= V \cos\theta(t),\\
\dot{y}(t) &= V \sin\theta(t),
\end{aligned}
\end{equation}
where the dot denotes a time derivative. The orientation angle evolves according to
\begin{equation}
\dot{\theta}(t) = \sqrt{2 D_R} \, \xi(t),
\label{eq:SDE_theta}
\end{equation}
where $D_R$ is the rotational diffusion coefficient and $\xi(t)$ is a zero-mean Gaussian white noise with delta-correlation $\langle \xi(t_1) \xi(t_2) \rangle = \delta(t_1 - t_2)$. 
This classical ABM captures persistent stochastic motion, but it bears no explicit representation of the periodic head motion associated with flagellar beating. We therefore propose a minimal extension in which the swimming direction is periodically modulated with oscillation frequency $\Omega$. 
A periodic angular modulation parameterized by a single oscillation amplitude and frequency constitutes the minimal extension of ABM that captures the cyclic directional changes observed in sperm trajectories. Furthermore, this modification preserves the simplicity and analytical tractability of the original framework.
Incorporating this periodic angular modulation leads to the oscillatory active Brownian motion (OABM) model,
\begin{equation}
\begin{aligned}
\dot{x}(t) &= V \cos \big[ \theta(t) + \psi(t) \big],\\
\dot{y}(t) &= V \sin \, \big[ \theta(t) + \psi(t) \big],
\end{aligned}
\label{eq:OABM}
\end{equation}
with $\psi(t)$ defined in equation~\eqref{eq:psi_def}. For a single trajectory, $V$, $A$, $\Omega$, $\phi$, and $D_R$ are fixed parameters (Table~\ref{tab:parameters}). At the population level, variability between cells is included, given empirical distributions $p(V)$, $p(A)$, $p(D_R)$, and $p(\Omega)$. In this way, the model remains minimal at the single-cell level while it allows cell-to-cell heterogeneity across the population. Whether these parameters should be sampled independently is discussed below.
The phase $\phi$ accounts for the fact that different sperm cells are observed at different points of their oscillatory cycle. For an ensemble of cells,  $\phi$ can be treated as being uniformly distributed in the interval $[0,2\pi)$.

In the experimental trajectories considered here, the observation time $T$ is typically long compared to the oscillation period, i.e.,
\begin{equation}
\frac{2\pi}{\Omega} \ll T .
\end{equation} 
In this regime, a single trajectory samples many beat cycles, but it does not necessarily sample the slow rotational decorrelation process. Therefore, averages over the oscillatory modulation can be interpreted as cycle averages, whereas $D_R$ should be handled cautiously. In particular, for individual trajectories, $D_R$ is better regarded as an effective directional relaxation rate rather than as a direct measurement of rotational diffusion.

\begin{table}[ht] 
\centering 
\begin{tabular}{lll} 
\hline 
Symbol & Description & Units \\ 
\hline 
$V$ & velocity & $\mu$m/s \\ 
$A$ & angular amplitude of oscillations & 1 \\
$\Omega$ & oscillation frequency & 1/s \\
$\phi$ & oscillation phase at $t=0$ & 1 \\
$D_R$ & rotational diffusion coefficient & 1/s \\ 
\hline 
\end{tabular} 
\caption{Parameters used in the OABM model and their definitions.} \label{tab:parameters} 
\end{table}


\section{Materials and Methods}


\subsection{Human sperm collection}
Experimental procedures involving human sperm were conducted in accordance with institutional ethical guidelines, reviewed and approved by the Institutional Review Board of Colorado State University, Colorado (IRB 7208).
Semen samples used in the study were provided by healthy donors, under written informed consent. Motile sperm were selected by the swim-up procedure \cite{Baldi2022} using (non-capacitating) modified human tubal fluid medium with gentamicin-HEPES (mHTF; FUJIFILM Irvine Scientific, Cat\#90126) supplemented with 0.5\% bovine serum albumin (BSA; Sigma Aldrich, Cat \#A7906).  1 mL of warm mHTF medium supplemented with 5 mg/mL BSA was placed in a 15-mL sterile conical tube and 0.5-1 mL of semen was gently layered beneath it without mixing. The tube was incubated for 1 h at a 45$^\circ$ angle at 37 $^\circ$C. After this time, the upper layer containing motile sperm was recovered and diluted with 1-2 mL of the same medium. The sample was centrifuged at 400$\times$g for 5 min, the supernatant was discarded and the sperm pellet was resuspended in fresh warm medium. Sperm concentration was estimated using a Makler chamber and adjusted to 5-10$\times 10^6$ cells/mL. 


\subsection{Induction of sperm hyperactivation}
Hyperactivating condition was induced by pharmacological elevation of intracellular cAMP using the analog 8-Br-cAMP (1 mM in mHTF supplemented with 0.5\% BSA at 37~$^\circ\mathrm{C}$) for 3 h prior to imaging) \cite{suarez2008control}.


\subsection{Image acquisition}

Sperm suspensions (10 µL) were placed on a microscope slide with an 18×18 mm glass coverslip covering them, resulting in a chamber depth of 30~$\mu$m. Negative phase-contrast videos were acquired using a Nikon Eclipse E200 microscope with a 10×/0.25 Ph1 BM objective and a Basler acA780-75gc camera at 60 fps. The pixel size was 0.83 $\mu$m. Each recording consisted of 240 frames (4 s). All samples were imaged at 37~$^\circ\mathrm{C}$ using a temperature-controlled stage. 

\subsection{Sperm tracking}
Trajectories were extracted using TrackMate \cite{Ershov2022trackmate}. Sperm head detection was performed using the Laplacian-of-Gaussian detector, with blob diameter set to 5 pixels and threshold set to 10. At our effective pixel size of $0.83~\mu$m, this blob diameter corresponds to approximately $4.2~\mu$m, which matches the typical size of a sperm head in our images. The median filter and sub-pixel localization were enabled, and no additional filters were applied to the detected spots. We note that the quality threshold in TrackMate scales with the intensity and contrast of the detected spots and is therefore specific to our imaging conditions. The value of 10 was chosen empirically as the lowest threshold that reliably detected sperm heads while excluding debris and background noise. Identified particle positions were linked using the simple LAP (linear assignment problem) tracker. A maximum linking distance of 5.5 pixels, a gap-closing maximum distance of 16.5 pixels, and a maximum gap-closing frame gap of 3 were used.  The resulting tracks were exported as XML files and converted to ASCII files in MATLAB for subsequent processing and handling.


\subsection{Numerical simulations}

The orientation angle $\theta(t)$ is a random process whose evolution is given by equation~\eqref{eq:SDE_theta}. For numerical simulations with a discrete time step $\Delta t$, with $\theta_i = \theta(t_i)$ and $t_i = i \,\Delta t$,
\begin{equation}
\theta_{i+1} = \theta_i + \sqrt{2 D_R} \, \Delta W_i,
\end{equation}
where $\Delta W_i$ is a Gaussian random variable with zero mean and variance $\Delta t$. The remaining equations in the OABM are deterministic. The simulations were performed in Python 3.12.5. The simulation codes used in this work are available on GitHub: \href{https://github.com/zindrok/OABM}{https://github.com/zindrok/OABM}.

\section{Characterization of sperm dynamics}

We use classical random-walk metrics, including time-averaged mean-squared displacement (TAMSD), velocity autocorrelation function (VACF), and power spectral density (PSD), to characterize motility. These parameters are summarized in Table~\ref{tab:metrics}. For a continuous trajectory $\mathbf{r}(t)$, the TAMSD in a lag time $\Delta$ is
\begin{equation}
\overline{\delta^2(\Delta)} = \frac{1}{T-\Delta} \int_0^{T-\Delta} \big|\mathbf{r}(t+\Delta) - \mathbf{r}(t) \big|^2 \, \mathrm{d}t.
\label{eq:TAMSD_def}
\end{equation}
This quantity measures how the squared displacement scales with lag time and is widely employed to characterize random walks \cite{metzler2014anomalous,munoz2021objective}.
Beyond the mean squared displacement (MSD), the time averaged VACF for a lag-time $\tau$ is \cite{rehfeldt2023random,sabri2020elucidating}
\begin{equation}
C_{\bf v}(\tau) = \frac{1}{T - \tau} \int_0^{T-\tau}  \mathbf{v}(t + \tau) \cdot \mathbf{v}(t) \, \mathrm{d}t.
\label{eq:VACF_def}
\end{equation}
where the velocity at time $t$ is given by the change in the increments, 
\begin{equation}
\mathbf{v}(t)=\frac{[\mathbf{r}(t+\delta t)-\mathbf{r}(t)]}{\delta t}.    
\end{equation}
where the increments are defined in a time interval $\delta t$, i.e., $\Delta\mathbf{r}(t)=\mathbf{r}(t+\delta t)-\mathbf{r}(t)$. When dealing with experimental trajectories, the increment time is given as a multiple of the sampling time $\delta t=k \Delta t$. At zero lag time, $C_{\bf v}(0) $ is the time-averaged squared velocity. Thus, we normalize the VACF as
\begin{equation}
\widehat{C}_{\bf v}(\tau) = \frac{C_{\bf v}(\tau)}{C_{\bf v}(0)}.
\label{eq:VACF_norm}
\end{equation}
We compute the normal PSD from the transverse motion $Y(t)$ of the sperm head relative to its progressive path. The PSD of the transverse motion is then \cite{krapf2018power}
\begin{equation}
S_{\rm N}(f) = \frac{1}{T} \Bigg| \int_0^{T} Y(t) \, e^{-i2\pi ft} \, \mathrm{d}t \Bigg|^2.
\label{eq:PSD_def}
\end{equation}
The dominant peak of this spectrum gives the main oscillation frequency of the sperm-head motion.

\begin{figure}[hbt!]
\centering
\includegraphics[width=0.3\textwidth]{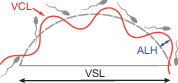}
\caption{\textbf{CASA metrics.} We consider the straight-line
velocity (VSL), curvilinear velocity (VCL), and amplitude of lateral head displacement (ALH).}
\label{fig:CASA}
\end{figure}

Next, we characterize sperm motility using the conventional terminology established in reproductive biology. Computer-assisted sperm analysis (CASA) provides a quantitative framework for assessing sperm motility through standardized kinematic parameters \cite{Lu2014,Mortimer2015,alquezar2019opencasa}. 
CASA-based analysis is widely used in both research and clinical settings. Furthermore, it constitutes the standard protocol recommended by the World Health Organization in its laboratory manual for the examination and processing of human semen \cite{world2021laboratory}. 
These metrics describe instantaneous velocity, net progression, and path geometry. Here, we consider the straight-line velocity (VSL), curvilinear velocity (VCL), and amplitude of lateral head displacement (ALH), as illustrated in Fig.~\ref{fig:CASA} and summarized in Table~\ref{tab:metrics}.

For each cell, we consider a discrete trajectory $\mathbf{r}(t_i)$ recorded at times $t_i=i\Delta t$, with $i=0,\dots,N$ and inter-frame time $\Delta t$. We compute two standard velocity measures: VSL and VCL (Fig.~\ref{fig:CASA}). VSL is the net displacement divided by the observation time,
\begin{equation}
\mathrm{VSL} = \frac{ \big| \mathbf{r}(t_N) - \mathbf{r}(t_0) \big| }{N \, \Delta t}.
\label{eq:VSL_def}
\end{equation}
VCL is the total trajectory length divided by the observation time,
\begin{equation}
\mathrm{VCL} = \frac{\displaystyle \sum_{i=0}^{N-1} \big| \mathbf{r}(t_{i+1}) - \mathbf{r}(t_i) \big| }{N\, \Delta t}.
\label{eq:VCL_def}
\end{equation}

To quantify the oscillatory motion of the sperm head, we compute the signed lateral displacement of the sperm head as follows. We define a smooth trajectory by employing a sliding-window moving average. For a window of size $t_w=w\Delta t$, the smooth trajectory is given by
\begin{equation}
\hat{\mathbf{r}}(t_i) = \frac{1}{w} \sum_{k=i}^{i+w-1} \mathbf{r}(t_k),
\label{eq:avrg_path}
\end{equation}
with $i=0,\dots,M$ and $M=N-w+1$. The deviation of the recorded position from this average path is $\Delta \mathbf{r}_i = \mathbf{r}(t_i) - \hat{\mathbf{r}}(t_i)$. We then compute the local unit tangent vector along the smooth path as $\mathbf{t}_i = \hat{\mathbf{v}}_i/|\hat{\mathbf{v}}_i|$, where $\hat{\mathbf{v}}_i = \big[\hat{\mathbf{r}}(t_{i+1}) - \hat{\mathbf{r}}(t_i) \big] / \Delta t$. The corresponding normal direction is obtained by rotating the unit tangent vector by $\pi/2$, $\mathbf{n}_i = \mathcal{R}_{\pi/2} \mathbf{t}_i$, where $\mathcal{R}_{\beta}$ denotes the rotation matrix for a rotation by angle $\beta$. The signed lateral displacement of the sperm head is then defined as $Y_i = \Delta \mathbf{r}_i \cdot \mathbf{n}_i$.
Positive and negative values of $Y_i$ correspond to the sperm head being on opposite sides of the average path. ALH (Fig.~\ref{fig:CASA}) is computed from $\ell_i = |Y_i|$, corresponding to the center-to-peak lateral amplitude. In this work, we use a window size $w = 8$.

The MATLAB codes to perform the analysis are available on GitHub: \href{https://github.com/zindrok/OABM}{https://github.com/zindrok/OABM}.

\begin{table}[ht] 
\centering 
\begin{tabular}{llll} 
\hline 
Acronym & Description & Equation & Symbol \\ 
\hline 
TAMSD & time-averaged mean squared displacement & \ref{eq:TAMSD_def} & $\overline{\delta^2(\Delta)}$ \\ 
VACF & velocity autocorrelation function & \ref{eq:VACF_def} & $C_{\bf v}(\tau)$ \\
PSD & power spectral density normal to the smooth path & \ref{eq:PSD_def} & $S_{\rm N}(f)$ \\
VSL & straight line velocity & \ref{eq:VSL_def} & VSL \\
VCL & curvilinear velocity & \ref{eq:VCL_def} & $V$ \\
ALH & amplitude of lateral head displacements & & $Y_{\rm max}$ \\ 
\hline 
\end{tabular} 
\caption{Metrics employed to characterize sperm dynamics.} 
\label{tab:metrics} 
\end{table}


\section{Results}


\subsection{Random walk metrics of OABM}

One advantage of OABM is that experimentally relevant observables can be obtained analytically. We first consider a single swimmer with fixed parameters $A$, $V$, $D_R$, $\phi$, and $\Omega$. Population-level heterogeneity will be later incorporated. 

The TAMSD is derived in appendix~\ref{app:TAMSD} for lag times shorter than the characteristic persistence time of the orientation. In this regime, the time is short compared with the slow directional decorrelation time, $\Delta\ll D_R^{-1}$,  while the trajectory may still contain many oscillation periods, and it is the more relevant to our experimental observations. Namely, 
\begin{equation}
\overline{\delta^2(\Delta)} \simeq 2 V^2 \int_0^\Delta \mathrm{d}\tau \, (\Delta-\tau)
J_0\bigg( 2A\sin\frac{\Omega \tau}{2} \bigg),\qquad  \Delta \ll D_R^{-1},
\label{eq:TAMSD_full}
\end{equation}
where the overbar denotes a temporal average and $J_0$ is the Bessel function of the first kind of order zero. We found a temporal crossover that is
controlled by the oscillatory time scale $\Omega^{-1}$. In the short and intermediate regimes, the TAMSD exhibits two asymptotes,
\begin{equation}
\overline{\delta^2(\Delta)} \sim 
     \begin{cases}
       V^2 \, \Delta^2, &  \Delta \ll \Omega^{-1} \qquad {\rm (short\ times)}, \\
       V^2 J^2_0(A)\, \Delta^2, & \Omega^{-1}\ll\Delta\ll D_R^{-1}\qquad {\rm (intermediate\ times)}, 
     \end{cases}
\label{eq:TAMSD}
\end{equation}
At short lag times, the motion is ballistic with velocity $V$. At lag times longer than the oscillation period but still shorter than the directional decorrelation time, the fast oscillatory component is averaged out. The motion remains ballistic, but with an effective velocity reduced by a factor $J_0(A)$, $V_{\rm eff}=VJ_0(A)$.
Interestingly, the ensemble-averaged MSD (EA-MSD) and the ensemble average of the TAMSD (EA-TAMSD) are virtually identical. As shown in figure \ref{fig:MSD} the EA-MSD $\langle \left|\mathbf{r}(\Delta)-\mathbf{r}(0)\right|^2\rangle$ and the EA-TAMSD $\langle \overline{\delta^2(\Delta)}\rangle$ obtained from numerical simulations collapse onto the same curve, with no discernible differences across the examined range of lag times.

The velocity autocorrelation function also contains a clear signature of the oscillatory angular modulation. In appendix~\ref{app:TAVACF}, we derive an approximate expression for the normalized time-averaged VACF. At short lag times, $\tau \ll D_R^{-1}$ yields
\begin{equation}
\widehat{C}_{\bf v}(\tau) \approx (1 - D_R \tau) \, J_0 \Biggr[ 2A \sin \biggr( \frac{\Omega \tau}{2} \biggr) \Biggr].
\label{eq:VACF}
\end{equation}
The prefactor $1-D_R\tau$ represents the leading-order contribution of slow directional decorrelation for $D_R\tau\ll1$. 

Finally, the maximum transverse excursion, derived in appendix~\ref{app:Max}, is
\begin{equation}
Y_{\rm max} = \frac{\pi V}{2 \Omega} \mathbf{H}_0(A),
\label{eq:y_max}
\end{equation}
where $\mathbf{H}_0$ is the Struve function of order zero. This expression links the measurable transverse amplitude of the oscillatory trajectory to the angular oscillation amplitude $A$, the swimming velocity $V$, and the beat frequency $\Omega$.


\subsection{Relations between OABM parameters and single trajectory statistics}

OABM parameters ($V$, $A$, $\Omega$, and $D_R$) are directly linked to the trajectory statistics. Thus, by obtaining parameters such as TAMSD and VCL from a single realization, one can estimate the OABM parameters that generate such a trajectory. The swimming velocity is estimated from the curvilinear velocity, 
\begin{equation}
 V = {\rm VCL}.
\end{equation}
The oscillation frequency can be estimated from the principal frequency of the normal PSD,
\begin{equation}
\Omega = 2 \pi f_{\rm max}.
\label{eq:omega}
\end{equation}
The angular oscillation amplitude is found from a non-linear fit of the TAMSD (Eq~\eqref{eq:TAMSD_full}). Because $V$ and $\Omega$ are obtained independently, the fitting involves a single free parameter $A$. However, the estimation procedure introduces a nonlinear correlation between these parameters, particularly between $A$ and $V$. Finally, the rotational diffusion coefficient $D_R$ is obtained from the slow decay of the envelope of the normalized VACF. Namely, at long lag times, this envelope is approximated by $1-D_R \tau$ (Eq~\eqref{eq:VACF}).

\begin{figure*}[hbt!]
\centering
\includegraphics[width=0.9\textwidth]{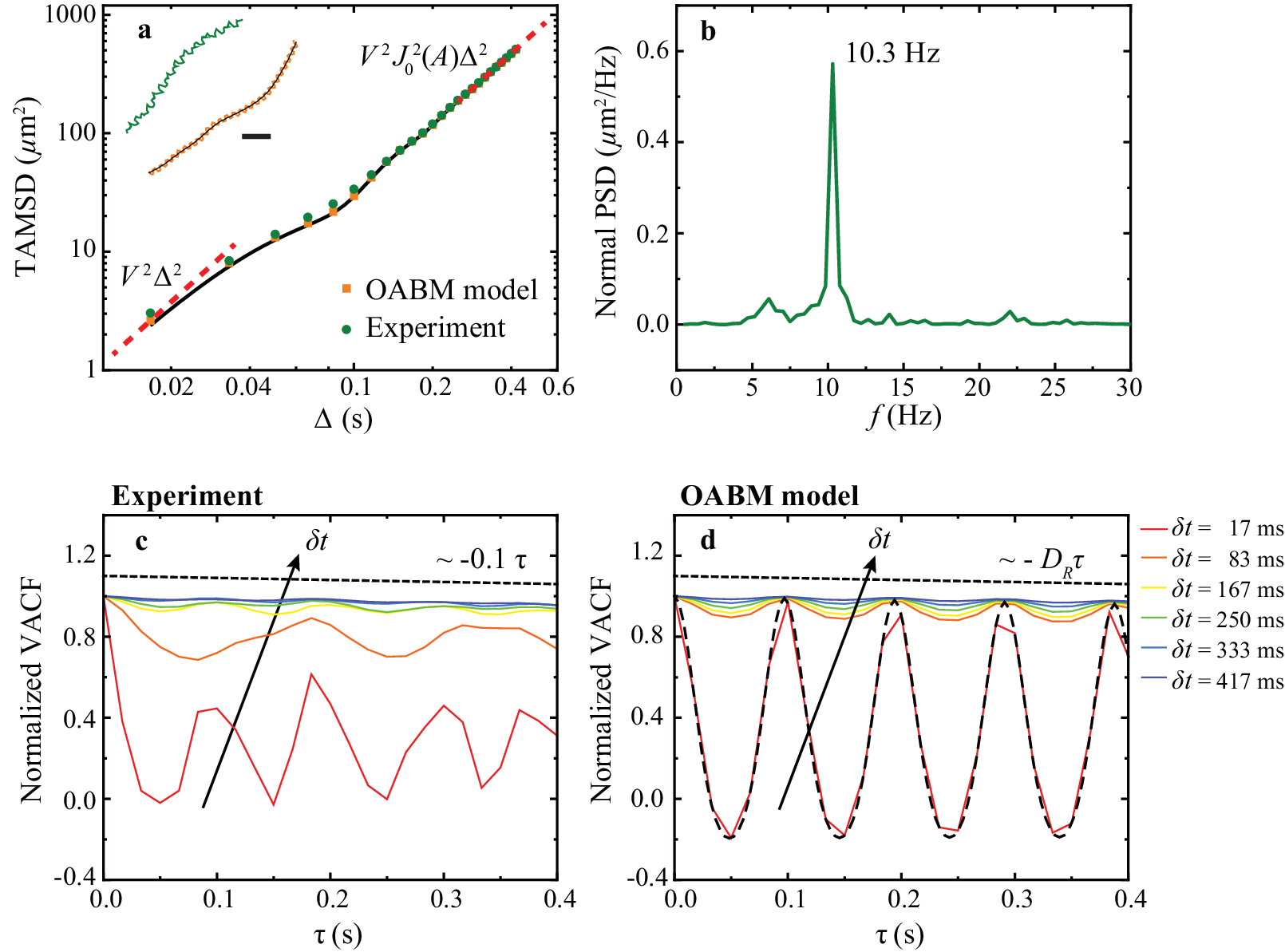}
\caption{\textbf{Comparison between an experimental sperm trajectory and its OABM counterpart.} 
(a)  ~Experimental and simulated TAMSD, showing the short-lag and intermediate-lag ballistic regimes. The solid black line shows the TAMSD fit to equation \eqref{eq:TAMSD_full} and the dashed red lines show the short-time and intermediate-time asymptotics (equation~\eqref{eq:TAMSD}). Inset: Representative experimental trajectory and synthetic OABM realization generated with parameters inferred from the experiment. Scale bar: $20~\mu\mathrm{m}$. 
(b)~Normal PSD of the experimental trajectory, with a dominant peak at $f_{\rm max}=10.3~\mathrm{Hz}$.
(c)~Normalized VACF of the experimental trajectory for several values of $\delta t$. For small $\delta t$, the VACF retains the oscillatory structure of the sperm-head motion. As $\delta t$ increases, the velocity averages over the fast oscillations, and the VACF approaches the slow directional-decorrelation envelope; the line with slope $\sim -D_R$ is shown as a guide. 
(d)~Normalized VACF for six values of $\delta t$. For $\delta t = 17$ ms, the normalized VACF follows equation~\eqref{eq:VACF}, shown as a dashed line. As $\delta t$ increases, the velocity averages over the fast oscillatory motion, suppressing the oscillatory component of the VACF. For large $\delta t$, the VACF is dominated by the slow directional-decorrelation envelope. The dashed line with slope $\sim -D_R$ is shown as a guide for comparison.  
}
\label{fig:comparison}
\end{figure*}


\subsection{Experimental sperm trajectories}

We acquired phase-contrast videos of live human sperm swimming in non-capacitating HTF medium and tracked the sperm-head positions to obtain experimental trajectories. For illustration purposes, we evaluated one representative trajectory consisting of 128 frames sampled at 60 FPS, shown in green in the inset of figure~\ref{fig:comparison}a. From this trajectory, we computed standard motility and dynamical observables, including VCL, ALH, TAMSD, VACF, and normal PSD. These measurements were then used to infer the corresponding OABM parameters. 
The VCL gives $V = 97~\mu\mathrm{m/s}$. 
The TAMSD is shown in figure~\ref{fig:comparison}a, where the crossover from the short-lag ballistic regime, $\overline{\delta^2(\Delta)}\sim V^2\Delta^2$, to the oscillation-averaged ballistic regime, $\overline{\delta^2(\Delta)}\sim V^2J_0^2(A)\Delta^2$, is observed. 
By fitting the TAMSD to equation \eqref{eq:TAMSD_full}, we
obtained the angular oscillation amplitude $A = 1.42$. Next, the normal PSD is shown in figure~\ref{fig:comparison}b. The PSD has a dominant peak at $f_{\rm max}=10.3~\mathrm{Hz}$, which corresponds to $\Omega = 65~\mathrm{s}^{-1}$ (equation~\eqref{eq:omega}).

Figure~\ref{fig:comparison}c shows the normalized VACF computed from the increments obtained with different lag times $\delta t$. For small $\delta t$, the VACF retains marked oscillations associated with the periodic head motion. As $\delta t$ increases, these oscillations are progressively suppressed. For $\delta t>300$~ms, the VACF is dominated by the slow envelope $1-D_R\tau$ (equation~\eqref{eq:VACF}), from which we estimate $D_R=0.1~\mathrm{s}^{-1}$. 
Additionally, the maximum and mean lateral head amplitudes are $\mathrm{ALH}_{\rm max}=2.27~\mu\mathrm{m}$ and $\mathrm{ALH}_{\rm mean}=1.18~\mu\mathrm{m}$, respectively. Using equation~\eqref{eq:y_max}, the inferred OABM parameters correspond to a transverse excursion amplitude $Y_{\rm max}=1.68~\mu\mathrm{m}$, which lies between $\mathrm{ALH}_{\rm mean}$ and $\mathrm{ALH}_{\rm max}$. This agreement supports the consistency of the inferred OABM parameters with standard sperm-motility descriptors.

Using the estimated OABM parameters, we simulated a trajectory according to equation~\eqref{eq:OABM}. The simulated trajectory is also shown in the inset of figure~\ref{fig:comparison}a. The synthetic trajectory reproduces the main geometric features of the measured path. The TAMSD in figure~\ref{fig:comparison}a shows the same crossover from the short-lag ballistic regime to the oscillation-averaged ballistic regime, as the experimental trajectory. The TAMSD confirms that the model captures the main features of the sperm-head motion. The normalized VACF of the OABM realization is shown in figure~\ref{fig:comparison}d. While the noise in the VACF that appears in the analysis of the experimental trajectory is lost, the VACF of the OABM realization captures several relevant patterns, including the oscillations at short lag times and the progressive loss of oscillations at long times. The VACF of the simulated trajectory also approximates the envelope $1-D_R\tau$ (equation~\eqref{eq:VACF}) for lag times $\delta t>300$~ms.


\subsection{Heterogeneous sperm population}

\begin{figure*}[hbt!]
\centering
\includegraphics[width=0.8\textwidth]{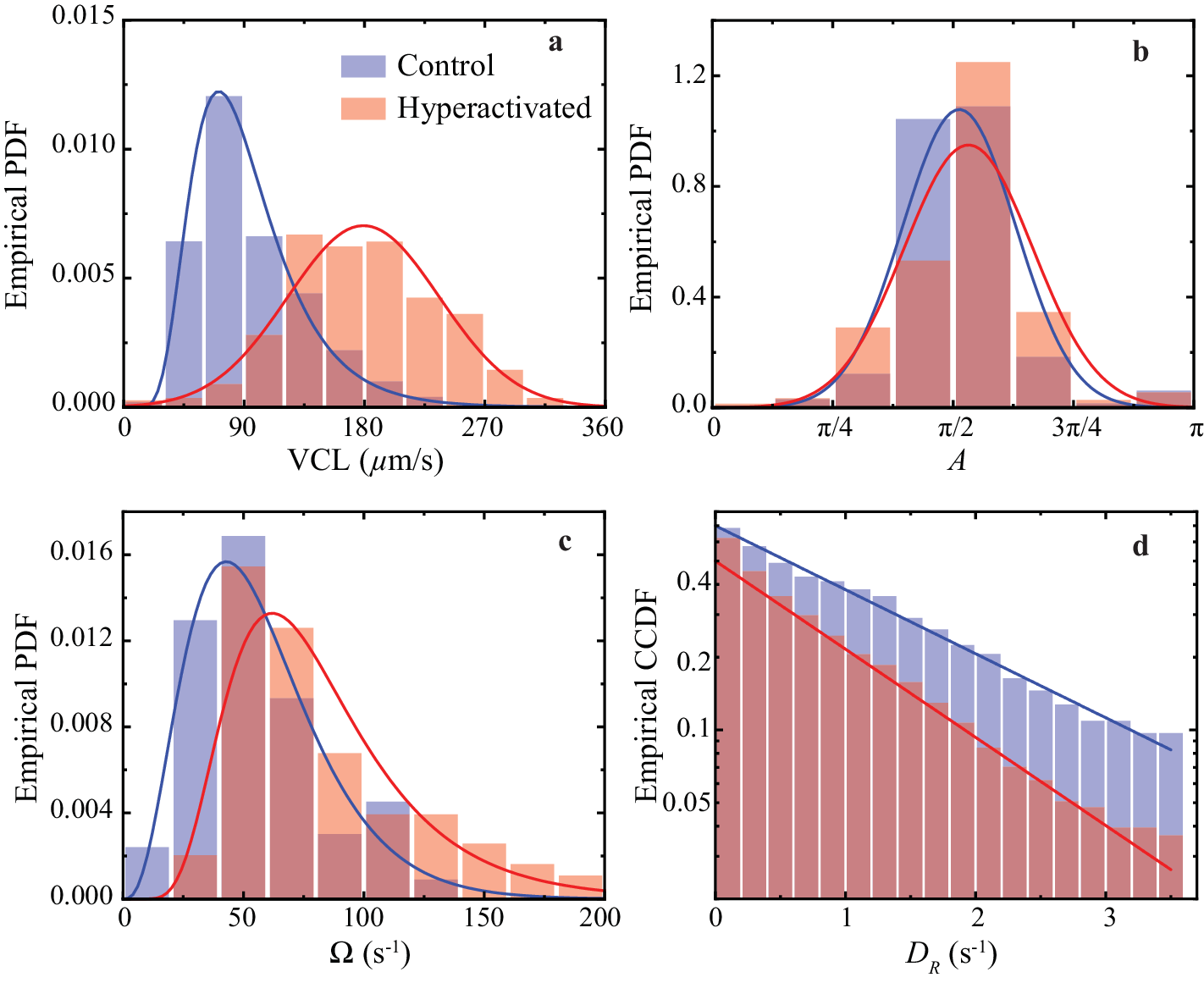}
\caption{\textbf{Empirical distributions of motility parameters obtained from $80$ control and $311$ sperm hyperactivated trajectories.} Solid lines show fits to the selected probability density functions.
(a)~Curvilinear velocity, VCL, fitted with a log-normal distribution for the control group and a Gaussian distribution for the hyperactivated group. 
(b)~Oscillatory amplitude $A$, fitted with a Gaussian distribution.
(c) Angular frequencies $\Omega$, fitted with a Gamma distribution for the control group and a log-normal for the hyperactivated group. 
(d) Complementary cumulative distribution function (CCDF) of the rotational diffusion coefficient $D_R$. The solid black lines correspond to exponential distributions.}
\label{fig:distribution}
\end{figure*}

We imaged and extracted trajectories of two populations of sperm cells: one in non-capacitating medium and another treated with 8-Br-cAMP to induce hyperactivation. We refer to the first as the control group and the second as the hyperactivated group. As commonly done in sperm motility analysis, we retained only trajectories with VSL $>$ 10 $\mu$m/s, thereby removing immotile trajectories. We obtained 178 control trajectories and 384 hyperactivated trajectories and 
extracted the OABM parameters from each. In 12 control trajectories and 18 hyperactivated trajectories, the fitting produced values of $A$ outside the physical range $(0,\pi)$. In these cases, the trajectories were discarded. Then, the distributions of the parameters were characterized using standard probability models. The results are shown in figure~\ref{fig:distribution}. The mathematical forms of the fitted distributions, along with details on the procedure used to select the best-fitting distribution are provided in appendix~\ref{app:best_fit}.
Figure~\ref{fig:distribution}a shows the distribution of swimming velocities $V$ for the control and hyperactivated groups. The control group is well described by a log-normal distribution, whereas the hyperactivated group is best described by a Gaussian distribution. The velocity distributions have means and standard deviations of $96.2 \pm 43.1~\mu\mathrm{m}/\mathrm{s}$ and $179.3 \pm 56.7~\mu\mathrm{m}/\mathrm{s}$ for control and hyperactivated groups, respectively. These values indicate an increase in the mean velocity and the spread of the distribution in the hyperactivated group, consistent with a key characteristic of these cells \cite{Mortimer1995Kine,Mortimer2000casa}. Figure~\ref{fig:distribution}b shows the distribution of oscillatory amplitudes, $A$. Both the control and hyperactivated group are described by Gaussian distributions. These distributions have means and standard deviations of $1.61 \pm 0.37$ and $1.67 \pm 0.42$ for the control and hyperactivated groups, respectively. Note that these distributions are concentrated around $A = \pi/2$. Figure~\ref{fig:distribution}c shows the distribution of the angular frequency $\Omega$. For the control group, the distributions of angular frequency is well approximated by Gamma distributions. The hyperactivated group, on the other hand, is best described by a log-normal distribution. Their means and standard deviations are $57.1 \pm 28.6~\mathrm{s}^{-1}$ and $82.8 \pm 38.3~\mathrm{s}^{-1}$ for control and hyperactivated groups, respectively. Finally, the rotational diffusion coefficient $D_R$ was estimated to follow an exponential distribution (figure~\ref{fig:distribution}d) with $\langle D_R \rangle = 1.64$ s$^{-1}$ for the control group and $\langle D_R \rangle = 1.19$ s$^{-1}$ for the hyperactivated group.

Supplementary figure~\ref{fig:Correlations} in appendix~\ref{app:Correlation} shows pairwise scatter plots of the fitted OABM parameters. It is observed that the velocity $V$ and angular amplitude $A$ are strongly correlated. Therefore, the marginal distributions $p(V)$ and $p(A)$ are not sufficient to describe the population. Instead, one needs the joint distribution $p(V,A)$. If an ensemble of OABM trajectories were simulated by sampling the above marginal distributions independently, the true trajectory-to-trajectory variability, in particular the spreading of the TAMSD, would not be correctly captured.
Estimating the full joint distribution $p(V,A)$ directly with limited data availability is challenging. Thus, we use a Gaussian copula as an alternative approach. We initially calculate the correlation coefficient $\rho$ of the two variables. We find $\rho_{\rm C} = 0.32$ for the control group and $\rho_{\rm H} = 0.60$ for the hyperactivated group. The Gaussian copula  is defined through the two-dimensional Gaussian distribution \cite{nelsen2006copulas}
\begin{equation}
p(G_1,G_2) =
\frac{1}{2\pi \sqrt{1-\rho^2}}
\exp \left[
-\frac{1}{2(1-\rho^2)}
\left(
G_1^2 - 2\rho G_1G_2 + G_2^2
\right)
\right],
\end{equation}
where $G_1$ and $G_2$ are standard normal random variables with correlation coefficient $\rho$, which as a first guess it equals the correlation coefficient of the variables we want to simulate. To generate a correlated pair $(V,A)$, we sample $(G_1,G_2)$ from this correlated Gaussian distribution and transform each Gaussian random variable into a uniform random variable via the standard normal cumulative distribution function $\Phi$, $u_1 = \Phi(G_1)$ and $u_2 = \Phi(G_2)$. Finally, we map these uniform variables to the desired marginals using the inverse cumulative distribution functions of $V$ and $A$, $V = F_V^{-1}(u_1)$ and $A = F_A^{-1}(u_2)$, where $F_V$ and $F_A$ are the cumulative distribution functions associated with the fitted marginal distributions of $V$ and $A$, respectively. This procedure preserves the fitted marginal distributions while introducing a correlation between $V$ and $A$. Because the resulting Pearson correlation between $V$ and $A$ does not necessarily match the input copula correlation exactly, we compute the correlation after transformation and recalibrate the copula parameter $\rho$ accordingly.

\begin{figure*}[hbt!]
\centering
\includegraphics[width=\textwidth]{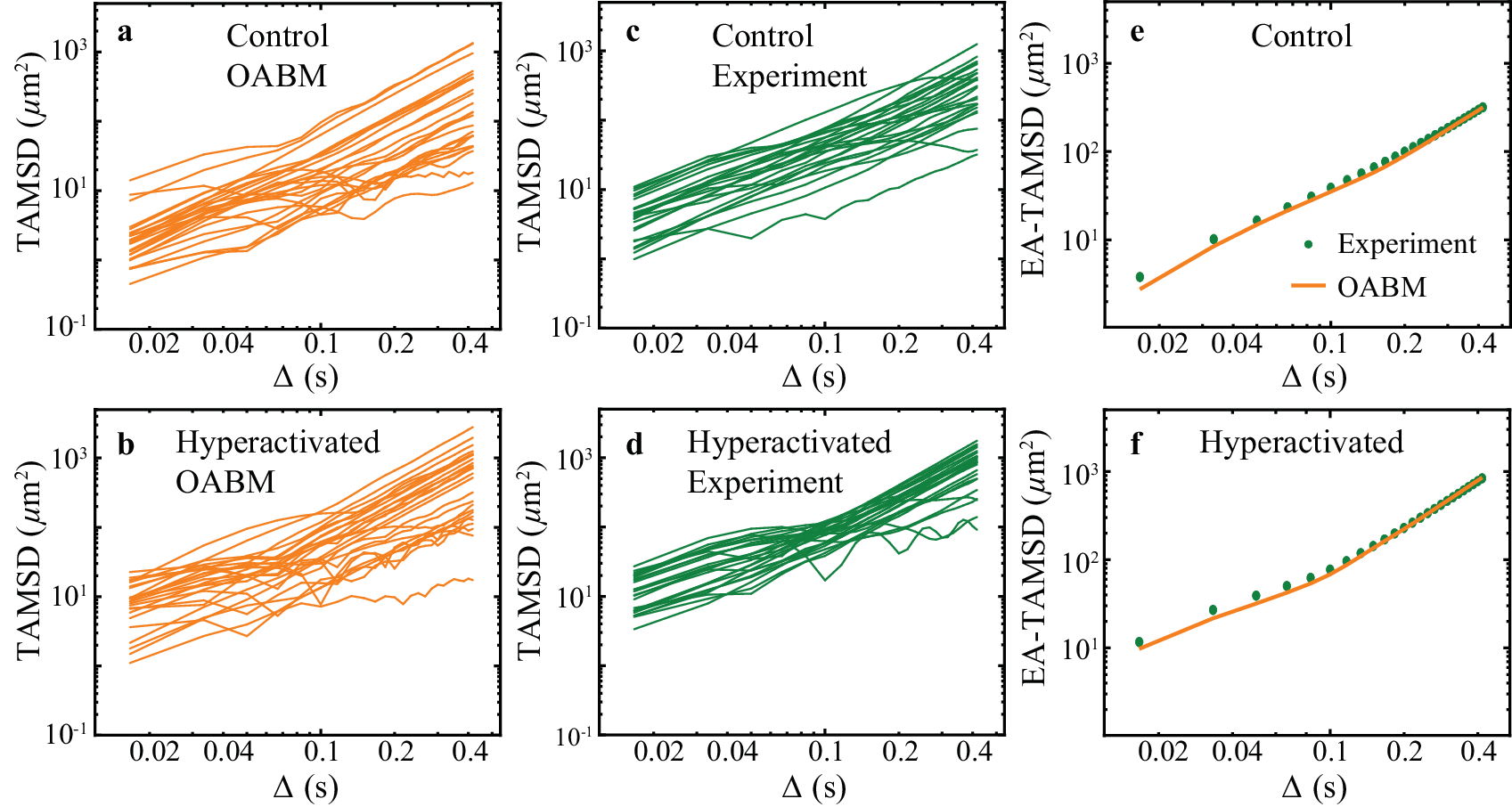}
\caption{\textbf{Population-level comparison between experimental sperm trajectories and OABM simulations.} (a)~TAMSDs of 25 randomly selected OABM realizations from the control group and (b)~after induction of hyperactivation. 
(c)~TAMSDs of 25 randomly selected experimental trajectories from the control group and (d)~after induction of hyperactivation. 
(e)~Ensemble-averaged TAMSDs for experiment and OABM in the control and (f)~hyperactivated groups.}
\label{fig:EATAMSD}
\end{figure*}

Using the above procedure, we generate an ensemble of 100 trajectories of 64 frames, with parameters drawn from
the measured distributions.
Figures~\ref{fig:EATAMSD}a and b show the TAMSDs of 25 randomly selected OABM trajectories for the control and hyperactivated groups, respectively, while figures~\ref{fig:EATAMSD}c and d show the corresponding experimental TAMSDs. Figures~\ref{fig:EATAMSD}e and f compare the ensemble-averaged TAMSDs obtained from experiment and from the OABM for the control and hyperactivated groups, respectively. The agreement between the experimental and OABM ensemble averages is excellent in both conditions, indicating that the model accurately captures both the population-level dynamics and the correct scaling. Furthermore, the comparable spread of the individual TAMSDs indicates that the joint sampling of $V$ and $A$ captures the main trajectory-to-trajectory variability observed experimentally.


\section{Discussion}

We developed oscillatory active Brownian motion (OABM) as a minimal model to describe the motion of human sperm. OABM reduces sperm swimming pattern into a small set of biologically interpretable parameters: swimming velocity, primary oscillatory frequency, oscillation amplitude, and directional persistence. By including a periodic angular drive, OABM captures both the progressive component of sperm swimming and the transverse oscillations associated with flagellar beating.
A central result of the model is the prediction of two ballistic regimes in the MSD. At very short lag times, the trajectory reflects the instantaneous swimming velocity. At longer times, after the fast oscillations have been averaged out, but before direction decorrelation, the trajectory remains ballistic but with a reduced effective velocity. This time dependence of the MSD provides a way to connect the angular oscillation amplitude $A$ to experimentally measured trajectory statistics. 
OABM also provides a useful way to interpret the velocity autocorrelation function. When the velocity is computed from increments at short time intervals, the VACF retains oscillations associated with periodic sperm-head motion. As the interval is increased, these oscillations are progressively averaged out, leaving the slower directional-decorrelation envelope. 
In practice, the VACF factorizes into a slow orientational relaxation term and a fast Bessel-function oscillatory term. The agreement between the experimental VACF and the model supports the interpretation that the observed trajectory oscillations arise from a periodic angular modulation superimposed on persistent active motion.

The model links the inferred angular oscillation amplitude to the transverse head excursion, a standard sperm-motility descriptor in both research and clinical applications \cite{world2021laboratory}. In this sense, all OABM parameters are linked to measurable quantities such as VCL, dominant sperm-head oscillation frequency, TAMSD, VACF, and lateral head displacement. Thus, the model addresses how flagellar beating oscillations contribute to sperm transport. This is important from a sperm biology perspective because while standard descriptors such as beat-cross frequency and amplitude of lateral head displacement are useful, they do not directly describe the physical components that generate sperm transport. This distinction is particularly relevant in the study of sperm physiological states. Hyperactivated sperm exhibit changes in lateral head displacement, sperm-head oscillations, swimming velocity, and directional persistence. By testing OABM in human sperm incubated under control conditions and after hyperactivation-inducing treatment, we show that these changes can be described across individual cells, treatments, and experimental conditions. In our data, the treated group showed changes that were consistent with hyperactivated-like motility, where sperm displayed stronger lateral motion, altered beat dynamics, and less regular trajectories.

An important aspect of OABM is that it is a kinematic model of sperm trajectories rather than a mechanical model of sperm propulsion. The periodic angular modulation represents the effect of flagellar beating on the swimming direction, without attempting to describe the detailed flagellar waveform or the surrounding flows. This level of description is justified because the dominant feature of the observed trajectories can be understood as persistent active motion modulated by a characteristic oscillatory component. By retaining only this leading-order oscillation, OABM captures the principal influence of flagellar beating on transport while remaining analytically tractable and requiring only a small number of experimentally accessible parameters.
The resulting synthetic trajectories reproduce important features of the experimental TAMSD distributions, supporting the ability of the OABM to capture the observed cell-to-cell heterogeneity. 

Broadly, OABM provides a compact active-matter framework for systems in which transport arises from the combination of persistent self-propulsion, stochastic reorientation, and internal cyclic driving. In this sense, from a physics perspective, the model produces simple but nontrivial motion, for which further investigation and extensions can be considered.
The simplicity of the model is both a strength and a limitation. OABM deliberately ignores details of the flagellar waveform, near-surface hydrodynamics, rolling motion, confinement, and interactions with boundaries or other cells. It also represents the beat-related oscillation through a single dominant angular frequency, even though experimental power spectra can contain weaker secondary peaks. Therefore, the model should not be interpreted as a full mechanical description of sperm flagellar propulsion. Instead, it provides a compact statistical description of how periodic sperm-head motion modifies coarse-grained transport. 
Despite the above limitations, OABM reproduced both individual human sperm trajectories and population-level behavior under control and hyperactivation-inducing conditions. The model explains how a periodic beat-related angular drive modifies persistent active motion, predicts experimentally accessible trajectory statistics, and links routine head-tracking measurements to physically interpretable parameters. This approach is useful for studying how capacitation-associated signaling pathways, pathological conditions in patients, pharmacological treatments, or heterogeneous environments, such as those encountered in the female reproductive tract, shape sperm transport.


\ack{APP and DK are grateful to Micha{\l} Balcerek and Agnieszka Wy{\l}oma\'nska for helpful discussions.}

\section*{Ethical statement} 

Experimental procedures involving human sperm were reviewed and approved by the Institutional Review Board of Colorado State University, Colorado, USA (IRB 7208). Semen samples were provided by healthy donors after written informed consent. All procedures were conducted in accordance with institutional ethical guidelines and applicable standards for research involving human participants.

\data{All data used in this study are publicly available in the Zenodo repository: \href{https://doi.org/10.5281/zenodo.20739357}{https://doi.org/10.5281/zenodo.20739357}.
The computational codes required to reproduce the simulations and analyses are available at \href{https://github.com/zindrok/OABM}{https://github.com/zindrok/OABM} and archived on Zenodo (same DOI).}

\section*{Conflict of interest} The authors declare no competing interests.

\section*{Author Contribution}

Adrian Pacheco-Pozo \orcid{0000-0003-2550-4566} \href{https://orcid.org/0000-0003-2550-4566}{0000-0003-2550-4566} \newline
Conceptualization (lead), Methodology (lead), Formal analysis (lead), Software (lead), Investigation (supporting), Visualization (lead), Writing - original draft (lead), Writing - review and editing (equal). \vspace{10pt}

\noindent Arturo Matamoros Volante \orcid{0000-0003-1750-3942} \href{https://orcid.org/0000-0003-1750-3942}{0000-0003-1750-3942} \newline
Conceptualization (supporting), Methodology (supporting), Formal analysis (supporting), Writing - review and editing (supporting). \vspace{10pt}

\noindent Pilar Ameijeiras \orcid{0000-0001-5486-6668} \href{https://orcid.org/0000-0001-5486-6668}{0000-0001-5486-6668} \newline 
Investigation (lead), Methodology (supporting), Data curation (supporting), Writing - review and editing (supporting). \vspace{10pt}

\noindent Mariano G. Buffone \orcid{0000-0002-7163-6482} \href{https://orcid.org/0000-0002-7163-6482}{0000-0002-7163-6482}\newline
Supervision (supporting), Resources (supporting), Methodology (supporting), Writing - review and editing (supporting). \vspace{10pt}

\noindent Diego Krapf \orcid{0000-0002-2833-5553} \href{https://orcid.org/0000-0002-2833-5553}{0000-0002-2833-5553}\newline 
Conceptualization (supporting), Methodology (supporting), Supervision (lead), Project administration (lead), Writing - original draft (supporting), Writing - review and editing (equal).\vspace{10pt}

\noindent All authors discussed the results, reviewed the manuscript, and approved the final version.


\appendix

\makeatletter
\renewcommand{\@seccntformat}[1]{Appendix~\csname the#1\endcsname.\quad}
\makeatother

\makeatletter
\@addtoreset{equation}{section}
\makeatother
\renewcommand{\theequation}{\Alph{section}.\arabic{equation}}

\setcounter{figure}{0}
\renewcommand{\thefigure}{S\arabic{figure}}

\setcounter{table}{0}
\renewcommand{\thetable}{S\arabic{table}}


\section{Time-averaged mean square displacement \label{app:TAMSD}}

For a single realization of the OABM, the TAMSD over a trajectory of duration $T$ at lag time $\Delta$ is
\begin{equation}
\overline{\delta^2(\Delta)} = \frac{1}{T-\Delta} \int_0^{T-\Delta} \big|\mathbf{r}(t+\Delta)-\mathbf{r}(t) \Big|^2 \, \mathrm{d}t .
\label{eq:ap_TAMSD}
\end{equation}
The OABM velocity is 
\begin{equation}
\dot{\mathbf r}(t) = V
\begin{bmatrix}
\cos [\theta(t)+\psi(t)] \\
\sin [\theta(t)+\psi(t)]
\end{bmatrix},
\end{equation}
where $\psi(t) = A \cos\big( \Omega t + \phi \big)$. The displacement over lag time $\Delta$ is therefore
\begin{equation}
\mathbf r(t+\Delta)-\mathbf r(t) = V\int_t^{t+\Delta}
\begin{bmatrix}
\cos [\theta(u)+\psi(u)] \\
\sin [\theta(u)+\psi(u)]
\end{bmatrix}
\,\mathrm{d}u .
\end{equation}
Substituting this expression into equation~\eqref{eq:ap_TAMSD} gives
\[
\overline{\delta^2(\Delta)} = \frac{V^2}{T-\Delta} \int_0^{T-\Delta} \left| \int_t^{t+\Delta}
\begin{bmatrix}
\cos [\theta(u)+\psi(u)] \\
\sin [\theta(u)+\psi(u)]
\end{bmatrix}
\,\mathrm{d}u \right|^2 \mathrm{d}t.
\]
Expanding the squared norm yields
\begin{equation}
\overline{\delta^2(\Delta)} = \frac{V^2}{T-\Delta} \int_0^{T-\Delta} \mathrm{d}t \int_t^{t+\Delta} \mathrm{d}u \int_t^{t+\Delta} \mathrm{d}u' \, \cos \big[\Theta(u)-\Theta(u')\big],
\end{equation}
where $\Theta(u)=\theta(u)+\psi(u)$. Using the changes of variables $u=t+s_1$ and $u'=t+s_2$, we obtain
\begin{equation}
\overline{\delta^2(\Delta)} = \frac{V^2}{T-\Delta} \int_0^{T-\Delta} \mathrm{d}t \int_0^\Delta \mathrm{d}s_1 \int_0^\Delta \mathrm{d}s_2\, \cos \Big[ \Theta(t+s_1)-\Theta(t+s_2) \Big].
\label{eq:ap_TAMSD_1}
\end{equation}
We now consider the regime in which the directional decorrelation time is much longer than the lag time, $\Delta \ll D_R^{-1}$. In this limit the orientation evolves slowly over a lag window, even though it does drift over the full trajectory. Consequently, the rotational angle $\theta(t)$ evaluated at $t+s$, with $0 \le s \le \Delta$, is essentially unchanged, i.e., $\theta(t+s) \approx \theta(t)$. Within this approximation the rotational contribution cancels in the angular difference of equation~\eqref{eq:ap_TAMSD_1}. Thus,
\begin{equation}
\Theta(t+s_1)-\Theta(t+s_2) = \psi(t+s_1)-\psi(t+s_2).
\end{equation}
Therefore,
\begin{equation}
\overline{\delta^2(\Delta)} = \frac{V^2}{T-\Delta} \int_0^{T-\Delta} \mathrm{d}t \int_0^\Delta \mathrm{d}s_1 \int_0^\Delta \mathrm{d}s_2 \, \cos\Big[ A\cos\big[\Omega(t+s_1)+\phi\big] - A\cos\big[\Omega(t+s_2)+\phi\big] \Big],
\end{equation}
where we have substituted the expression for $\psi(t)$. Using the trigonometric identity
\begin{equation}
\cos x-\cos y = -2\sin\left(\frac{x+y}{2}\right) \sin\left(\frac{x-y}{2}\right),
\end{equation}
gives
\begin{equation}
\psi(t+s_1)-\psi(t+s_2) = -2A \sin\bigg[ \Omega t+\frac{\Omega(s_1+s_2)}{2}+\phi \bigg] \sin\bigg[ \frac{\Omega(s_1-s_2)}{2} \bigg].
\label{eq:ap_diff_angles}
\end{equation}
Because the cosine is an even function,
\begin{equation}
\cos\big[ \psi(t+s_1)-\psi(t+s_2) \big] = \cos\Bigg[2A \sin\bigg[ \Omega t+\frac{\Omega(s_1+s_2)}{2}+\phi \bigg] \sin\bigg[ \frac{\Omega(s_1-s_2)}{2} \bigg] \Bigg].
\end{equation}
Thus,
\begin{equation}
\overline{\delta^2(\Delta)} = \frac{V^2}{T-\Delta} \int_0^\Delta \mathrm{d}s_1
\int_0^\Delta \mathrm{d}s_2 \int_0^{T-\Delta} \mathrm{d}t \, \cos\Bigg[2A \sin\bigg[ \Omega t+\frac{\Omega(s_1+s_2)}{2}+\phi \bigg] \sin\bigg[ \frac{\Omega(s_1-s_2)}{2} \bigg] \Bigg].
\label{eq:ap_TAMSD_three_int}
\end{equation}
We now concentrate on the inner time integral,
\begin{equation}
I = \frac{1}{T-\Delta} \int_0^{T-\Delta} \cos\Bigg[2A \sin\bigg[ \Omega t+\frac{\Omega(s_1+s_2)}{2}+\phi \bigg] \sin\bigg[ \frac{\Omega(s_1-s_2)}{2} \bigg] \Bigg] \, \mathrm{d}t .
\end{equation}
By defining $z = 2A\sin\left[ \Omega(s_1-s_2)/2 \right]$ and $\alpha =[ \Omega(s_1+s_2)/2 ]+\phi$, we can write this as
\begin{equation}
I = \frac{1}{T-\Delta} \int_0^{T-\Delta} \cos\left[z \sin\left( \Omega t+\alpha\right)\right] \, \mathrm{d}t.
\end{equation}
Taking the change of variable $x = \Omega t + \alpha$, we obtain
\begin{equation}
I = \frac{1}{L} \int_\alpha^{\alpha+L} \cos(z\sin x) \,  \mathrm{d}x,
\end{equation}
with $L = \Omega (T-\Delta)$. We now assume that $L$ contains many complete oscillations. Writing $L = 2 \pi N + r$, with $N$ the number of complete oscillations and $0\leq r < 2\pi$, we split the above integral as
\begin{equation}
I = \frac{1}{L} \int_\alpha^{\alpha+ 2\pi N } \cos(z\sin x) \,  \mathrm{d}x + \frac{1}{L} \int_{\alpha+2\pi N}^{\alpha + 2 \pi N + r} \cos(z\sin x) \,  \mathrm{d}x.
\end{equation}
Since the cosine function is bounded in absolute value by unity, the second term is bounded by $r/L$ and is therefore of order $\mathcal{O}(1/L)$. Moreover, since the cosine function is $2\pi$-periodic, the first integral reduces to $N$ identical integrals over one period, Therefore
\begin{equation}
I = \frac{N}{L} \int_0^{2\pi} \cos(z\sin x) \,  \mathrm{d}x + \mathcal{O}\bigg(\frac{1}{L} \bigg).
\end{equation}
In the case where $L \gg 2\pi$, meaning that it contains a great number of complete oscillations, we can use $L/N \approx 2\pi$, and approximate this integral by
\begin{equation}
I \approx \frac{1}{2\pi} \int_0^{2\pi} \cos\big(z\sin u\big)\,\mathrm{d}u = J_0(z),
\end{equation}
which is the integral representation of the Bessel function of order zero \cite{NIST:DLMF}. Therefore, returning to the equation~\eqref{eq:ap_TAMSD_three_int}, we have
\begin{equation}
\overline{\delta^2(\Delta)} \simeq V^2 \int_0^\Delta \mathrm{d}s_1 \int_0^\Delta \mathrm{d}s_2 \, J_0 \Bigg[ 2A\sin \bigg[\frac{\Omega(s_1-s_2)}{2} \bigg] \Bigg].
\end{equation}
Since the integrand depends only on the difference $s_1-s_2$, this double integral can be reduced to 
\begin{equation}
\overline{\delta^2(\Delta)} \simeq 2 V^2 \int_0^\Delta \mathrm{d}\tau \, (\Delta-\tau)
J_0\bigg( 2A\sin\frac{\Omega \tau}{2} \bigg).
\label{eq:ap_TAMSD_2}
\end{equation}
This equation describes the TAMSD in the regime where rotational decorrelation is negligible over the trajectory duration, but the oscillatory modulation is fully sampled.

For very short lag times compare to the period of oscillation, $\tau \ll \Omega^{-1}$, the relevant small parameter is the argument of the Bessel function, $z(\tau)=2A\sin (\Omega\tau/2)$. Thus, the expansion below requires $z(\tau)\ll 1$. For $\Omega\tau\ll 1$, this condition becomes $A\Omega\tau\ll 1$. Therefore, for short lags satisfying $A\Omega\Delta\ll 1$, we expand
\begin{equation}
J_0 \left( 2A \sin \frac{\Omega \tau}{2} \right) = 1 - \frac{A^2 \Omega^2}{4}\,\tau^2 + \mathcal{O}(\tau^4).
\end{equation}
Substituting into equation~\eqref{eq:ap_TAMSD_2} gives
\begin{equation}
\overline{\delta^2(\Delta)}  = V^2 \bigg[ \Delta^2  - \frac{A^2 \Omega^2}{24}\Delta^4 + \mathcal{O}\big(\Delta^6\big) \bigg].
\end{equation}
Thus, at very short times,
\begin{equation}
\overline{\delta^2(\Delta)} \simeq V^2 \, \Delta^2.
\end{equation}
This is the usual ballistic regime, governed by the instantaneous swimming velocity $V$.

We now consider the intermediate window $\Omega^{-1}\ll \Delta\ll D_R^{-1}$. In this window, rotational decorrelation is still negligible, but the lag time is long compared with the oscillation period. Equation~\eqref{eq:ap_TAMSD_1} therefore applies. Applying the Jacobi-Anger expansion \cite{NIST:DLMF} to the integrand in equation~\eqref{eq:ap_TAMSD_2}, and integrating term-by-term, gives the exact decomposition
\begin{equation}
\overline{\delta^2(\Delta)} = V^2 J_0^2(A) \, \Delta^2 + \frac{ 4 V^2}{\Omega^2}\sum_{n=1}^\infty \frac{J_n^2(A)}{n^2} \, \big[1-\cos(n\Omega \Delta)\big].
\end{equation}
The first term grows quadratically with lag time, whereas the second term is bounded and oscillatory. Indeed, each term in the sum is of order $(n\Omega)^{-2}$, so the sum produces only finite oscillatory corrections and a constant offset. Therefore, the dominant intermediate-lag behavior is
\begin{equation}
\overline{\delta^2(\Delta)} \simeq V^2\,J_0^2(A)\,\Delta^2.
\end{equation}
Hence, the TAMSD displays two ballistic regimes:
\begin{equation}
\overline{\delta^2(\Delta)} \sim 
\begin{cases}
V^2\Delta^2,
& \Delta\ll \Omega^{-1}, \\[6pt]
V^2J_0^2(A)\Delta^2,
& \Omega^{-1}\ll \Delta\ll D_R^{-1}.
\end{cases}
\end{equation}
The first regime reflects motion at the instantaneous swimming velocity $V$. The second regime reflects motion at the oscillation-averaged effective velocity $V J_0(A)$.

\begin{figure*}[hbt!]
\centering
\includegraphics[width=0.5\textwidth]{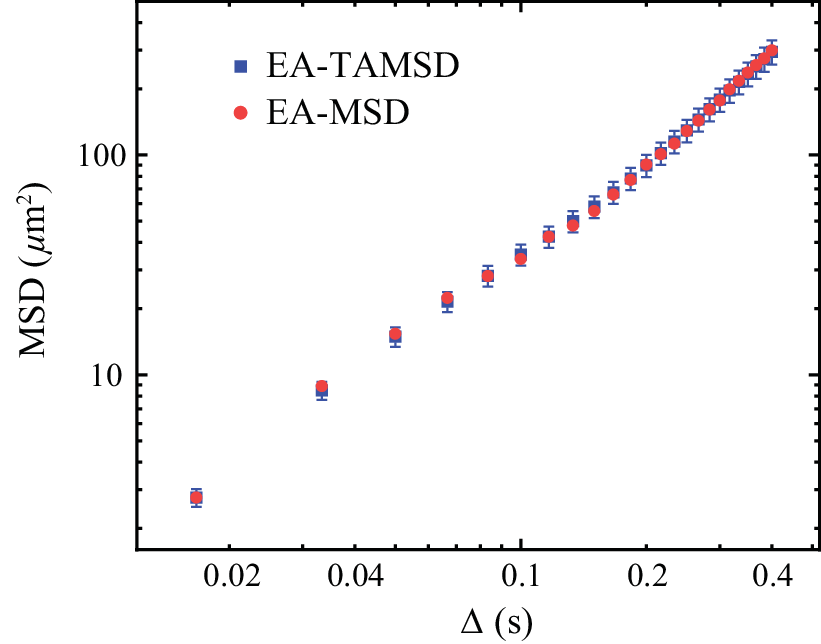}
\caption{\textbf{Comparison between the ensemble-average of the TAMSD (EA-TAMSD) and the ensemble-average MSD (EA-MSD).} Both observables were computed from 100 simulated trajectories of 64 frames each, using the parameters values obtained from the control group in the main text.}
\label{fig:MSD}
\end{figure*}


\section{Temporal-average velocity autocorrelation function \label{app:TAVACF}}

We consider the OABM velocity
\begin{equation}
{\bf v}(t) = V
\begin{pmatrix}
\cos[\theta(t)+\psi(t)]\\
\sin[\theta(t)+\psi(t)]
\end{pmatrix},
\label{eq:ap_vel}
\end{equation}
where $\psi(t)=A\cos(\Omega t + \phi)$. The ensemble-average velocity autocorrelation function of a single trajectory is
\begin{equation}
C_{\bf v}(\tau) = \frac{1}{T-\tau} \int_0^{T-\tau} \textbf{v}(t) \cdot \textbf{v}(t+\tau) \, \textrm{d}t.
\end{equation}
Using equation~\eqref{eq:ap_vel},
\begin{equation}
\textbf{v}(t) \cdot \textbf{v}(t+\tau) = V^2 \cos \big[ \theta(t+\tau) - \theta(t) + \psi(t+\tau) - \psi(t) \big].
\end{equation}
We define the angular increments $\Delta\theta_\tau(t) = \theta(t+\tau)-\theta(t)$ and $\Delta\psi_\tau(t) = \psi(t+\tau)-\psi(t)$. Then,
\begin{equation}
C_{\bf v}(\tau) = \frac{V^2}{T-\tau} \int_0^{T-\tau} \cos\big[ \Delta\psi_\tau(t) + \Delta\theta_\tau(t) \big] \mathrm{d}t .
\label{eq:ap_TA-VACF}
\end{equation}

We now focus on the short rotational-diffusion regime, $D_R \tau \ll 1$. In this regime, the rotational increment over the lag time $\tau$ is small, i.e., $\Delta\theta_r(t)\ll 1$. Expanding to second order in $\Delta\theta_r(t)$, gives
\begin{equation}
\cos\big[ \Delta\psi_\tau(t) + \Delta\theta_\tau(t) \big] \simeq \cos\Delta\psi_\tau(t) - \Delta\theta_\tau(t)\sin\Delta\psi_\tau(t) -
\frac{1}{2} \big[ \Delta\theta_\tau(t) \big]^2 \cos\Delta\psi_\tau(t).
\end{equation}
Substituting this expression into equation~\eqref{eq:ap_TA-VACF} gives
\begin{align}
C_{\bf v}(\tau) \simeq V^2 I_0(\tau,T) - V^2 I_1(\tau,T) - \frac{V^2}{2}I_2(\tau,T),
\label{eq:ap_TAVACF_1}
\end{align}
where
\begin{equation}
I_0(\tau,T) = \frac{1}{T-\tau} \int_0^{T-\tau} \cos\Delta\psi_\tau(t) \, \mathrm{d}t,
\end{equation}
\begin{equation}
I_1(\tau,T) = \frac{1}{T-\tau} \int_0^{T-\tau} \Delta\theta_\tau(t) \sin\Delta\psi_\tau(t) \, \mathrm{d}t,
\end{equation}
and
\begin{equation}
I_2(\tau,T) = \frac{1}{T-\tau} \int_0^{T-\tau} [\Delta\theta_\tau(t)]^2 \cos\Delta\psi_\tau(t) \, \mathrm{d}t.
\end{equation}

Using equation~\eqref{eq:ap_diff_angles} from~\ref{app:TAMSD}, we can write the  deterministic angular increment as 
\begin{equation}
\Delta\psi_\tau(t) = -B\sin(\Omega t+\phi'),
\end{equation}
with $B = 2A \sin (\Omega \tau /2 )$ and $\phi' = \Omega\tau /2 -\phi$.
For an observation window containing many oscillation periods, $T-\tau\gg 2\pi / \Omega$, the average over the deterministic oscillation can be replaced by a cycle average. Therefore,
\begin{equation}
I_0(\tau,T) \simeq \frac{1}{2\pi} \int_0^{2\pi} \cos \big( B\sin u \big) \, \mathrm{d}u = J_0(B).
\label{eq:ap_I0}
\end{equation}
The term $I_1$ is linear in the signed rotational increment. Since $\sin\Delta\psi_\tau(t)$ oscillates on the fast time scale $\Omega^{-1}$, while $\Delta\theta_\tau(t)$ varies on the slower rotational-diffusion time scale, this term gives a
subleading finite-time contribution over many oscillation periods. We therefore approximate 
\begin{equation}
I_1(\tau,T)\simeq 0.
\label{eq:ap_I1}
\end{equation}
For the quadratic contribution, we introduce the trajectory-level angular mean-square increment
\begin{equation}
Q_\theta(\tau) = \frac{1}{T-\tau} \int_0^{T-\tau} [\Delta\theta_\tau(t)]^2 \, \mathrm{d}t.
\end{equation}
Then, decompose the squared angular increment into its time average and a fluctuation, $[\Delta\theta_\tau(t)]^2 = Q_\theta(\tau) + \eta_\tau(t)$, with
\begin{equation}
\frac{1}{T-\tau}
\int_0^{T-\tau}\eta_\tau(t) \, \mathrm{d}t = 0.
\end{equation}
Substituting this decomposition into $I_2$ gives
\begin{equation}
I_2(\tau,T) = Q_\theta(\tau) \, \frac{1}{T-\tau} \int_0^{T-\tau} \cos\Delta\psi_\tau(t) \, \mathrm{d}t + \frac{1}{T-\tau} \int_0^{T-\tau} \eta_\tau(t) \, \cos\Delta\psi_\tau(t) \, \mathrm{d}t.
\end{equation}
The second term is a covariance-like correction between the slowly varying stochastic fluctuation $\eta_\tau(t)$ and the deterministic oscillatory factor $\cos\Delta\psi_\tau(t)$. When the oscillatory factor varies on a much shorter time scale than $\eta_\tau(t)$, or when the two factors are effectively uncorrelated over the averaging window, this correction is subleading. Thus, to leading order,
\begin{equation}
I_2(\tau,T) \simeq Q_\theta(\tau) \frac{1}{T-\tau} \int_0^{T-\tau} \cos\Delta\psi_\tau(t) \, \mathrm{d}t.
\end{equation}
Using equation~\eqref{eq:ap_I0}, $I_2(\tau,T) \simeq  Q_\theta(\tau) J_0(B)$. For angular Brownian motion, the short-lag mean-square angular increment is $Q_\theta(\tau) \simeq 2D_R\tau$ provided that $T \gg \tau$, so that sufficiently many lagged increments are sampled. Hence,
\begin{equation}
I_2(\tau,T) \simeq 2D_R\tau J_0(B).
\label{eq:ap_I2}
\end{equation}
Substituting equations~\eqref{eq:ap_I0},~\eqref{eq:ap_I1}, and~\eqref{eq:ap_I2} into equation~\eqref{eq:ap_TAVACF_1}, we obtain
\begin{equation}
C_{\bf v}(\tau) \simeq V^2(1-D_R\tau) J_0\left[ 2A\sin\left(\frac{\Omega\tau}{2}\right) \right].
\end{equation}
Since $C_{\bf v}(0) = V^2$, the normalized velocity autocorrelation function is
\begin{equation}
\widehat{C}_{\bf v}(\tau) = \frac{C_{\bf v}(\tau)}{C_{\bf v}(0)} \simeq (1-D_R\tau) J_0\left[ 2A\sin\left(\frac{\Omega\tau}{2}\right) \right].
\label{eq:ap_normal}
\end{equation}
This finite-time approximation is valid for a single trajectory under the conditions $D_R \tau \ll 1$, $T - \tau \gg \Omega^{-1}$, $T\gg \tau$, and when the trajectory-level angular mean-square increment satisfies $Q_\theta(\tau) \simeq 2D_R\tau$. Equation~\eqref{eq:ap_normal} shows that the normalized VACF factorizes, to leading order, into a slow orientational decorrelation envelope and a fast oscillatory contribution. The envelope is $1-D_R \tau$, while the oscillatory component is controlled by the Bessel factor
\begin{equation}
J_0\left[ 2A\sin\left(\frac{\Omega\tau}{2}\right) \right].
\end{equation}


\section{Maximal oscillation \label{app:Max}}

To separate the rapid oscillatory dynamics from rotational diffusion, we use a reference frame attached to the swimmer’s instantaneous heading. We define the orthonormal basis
\begin{align}
\nonumber \mathbf{e}_\parallel(t) &= \big[\phantom{-}\cos\theta(t), \,\sin\theta(t)\big]^\top, \\
\mathbf{e}_\perp(t) &= \big[-\sin\theta(t),\,\cos\theta(t)\big]^\top,
\end{align}
and the rotation matrix 
\begin{equation}
\mathcal{E}(t)=[\mathbf e_\parallel(t)\ \mathbf e_\perp(t)].
\end{equation}
Projecting the laboratory-frame velocity onto this basis gives 
\begin{equation}
\begin{bmatrix}
v_\parallel(t)\\[2pt]
v_\perp(t)
\end{bmatrix}
=\mathcal{E}(t)^\top 
\begin{bmatrix}
\dot{x}(t)\\[2pt]
\dot{y}(t)
\end{bmatrix}.
\end{equation}
From the OABM, we know that
\begin{equation}
\begin{bmatrix}
\dot{x}(t)\\[2pt]
\dot{y}(t)
\end{bmatrix} = V
\begin{bmatrix}
\cos[\theta(t)+\psi(t)]\\[2pt]
\sin[\theta(t)+\psi(t)]
\end{bmatrix},
\end{equation}
so the projections are exactly
\begin{align}
\nonumber v_\parallel(t) & = V \cos\psi(t),\\
v_\perp(t) & = V \sin\psi(t). 
\end{align}

We now consider the transverse displacement in the body frame,
\begin{equation}
Y(t) = \int_{t_0}^t v_\perp(t') \, \mathrm{d}t' = V \int_{t_0}^t \sin \big[ A \cos (\Omega t' + \phi) \big] \mathrm{d}t',
\end{equation}
where $t_0$ is a reference time to be determined. Expanding the integrand using the Jacobi–Anger formula \cite{NIST:DLMF} yields
\begin{equation}
Y(t) = \frac{2V}{\Omega}\sum_{k=0}^{\infty} (-1)^k \frac{J_{2k+1}(A)}{2k+1} \sin \big[(2k{+}1)(\Omega t+\phi)\big]-C(t_0),
\end{equation}
where
\begin{equation}
C(t_0) = \frac{2V}{\Omega}\sum_{k=0}^{\infty} (-1)^k \frac{J_{2k+1}(A)}{2k+1} \sin \big[(2k{+}1)(\Omega t_0+\phi)\big].
\end{equation}
One could take $t_0 = 0$, for which the resulting $Y(t)$ would contain an additional constant offset that depends on $\phi$. This offset only changes the reference value of $Y$ and does not affect the oscillation amplitude. A more convenient choice for $t_0$ is $t_0 = - \phi / \Omega$, so that $\Omega t_0 + \phi = 0$. With this choice, every sine term in $C(t_0)$ vanishes and hence $C(t_0)=0$. Therefore,
\begin{equation}
Y(t) = \frac{2V}{\Omega}\sum_{k=0}^{\infty} (-1)^k \frac{J_{2k+1}(A)}{2k+1} \sin \big[(2k{+}1)(\Omega t+\phi)\big].
\end{equation}
This representation shows that $Y(t)$ contains only odd harmonics of the driving frequency $\Omega$. It also implies $\langle Y(t) \rangle_T = 0$ over one period $T = 2 \pi/\Omega$.

The extrema of $Y(t)$ are obtained from the condition
\begin{equation}
\dot{Y}(t) = v_\perp (t) = V \sin \big[ A \cos(\Omega t + \phi)\big] = 0
\end{equation}
For $0 \leq A < \pi$, the first positive maximum occurs at $\Omega t + \phi = \pi / 2$. Evaluating $Y(t)$ at the maximum gives
\begin{equation}
Y_{\rm max} = \frac{2 V}{\Omega} \sum_{k = 0}^{\infty} \frac{J_{2k+1}(A)}{2k+1}.
\end{equation}
 This result can be written in terms of the Struve function of order zero defined as
\begin{equation}
\mathbf{H}_0(x) = \frac{2}{\pi} \int_0^{\pi/2} \sin \big( x \cos t \big) \, \mathrm{d}t = \frac{4}{\pi} \sum_{k = 0}^{\infty} \frac{J_{2k+1}(x)}{2k+1},
\end{equation}
see equations~(11.5.1) and~(11.4.21) in Ref.~\cite{NIST:DLMF}. Hence,
\begin{equation}
Y_{\rm max} = \frac{\pi V}{2 \Omega} \mathbf{H}_0(A).
\end{equation}


\section{Best fit distribution \label{app:best_fit}}

To construct the population-level OABM simulations, we fitted empirical probability distributions to the model parameters extracted from the experimental trajectories. The goal of this procedure was to obtain a statistical description of cell-to-cell heterogeneity in the swimming velocity $V$, oscillatory amplitude $A$, the angular frequency $\Omega$, and the rotational diffusion coefficient $D_R$.

For the swimming velocity $V$, the angular frequency $\Omega$, and the oscillatory amplitude, we selected between four candidate distributions: Gamma, log-normal, Weibull, and Gaussian \cite{carlton2017proba}. These distributions were fitted separately for each experimental condition, control and hyperactivated. The candidate probability density functions were as follows. The Gamma distribution was written as
\begin{equation}
p(x) = \frac{\beta^\lambda}{\Gamma(\lambda)} x^{\lambda-1}e^{-\beta x}, \qquad x>0,
\end{equation}
where $\lambda$ is the shape parameter and $\beta$ is the rate parameter. The log-normal distribution was written as
\begin{equation}
p(x) = \frac{1}{x\sigma\sqrt{2\pi}} \exp\Bigg[-\frac{(\ln x-\mu)^2}{2\sigma^2}\Bigg], \qquad x>0,
\end{equation}
where $\mu$ and $\sigma$ are the mean and standard deviation of $\ln x$. The Weibull distribution was written as
\begin{equation}
p(x) = \frac{n}{\zeta} \left(\frac{x}{\zeta}\right)^{n-1} \exp\left[ -\left(\frac{x}{\zeta}\right)^n \right], \qquad x>0,
\end{equation}
where $\zeta$ is the scale parameter and $n$ is the shape parameter. The Gaussian distribution was written as
\begin{equation}
p(x) = \frac{1}{\sqrt{2\pi \sigma^2}} \exp \bigg[-\frac{(x - \mu)^2}{2\sigma^2} \bigg],
\end{equation}
where $\mu$ is the mean and $\sigma$ its standard deviation. Note that the first three families, gamma, log-normal, and Weibull, have support on $(0,\infty)$, whereas the Gaussian distribution has support on $(-\infty,\infty)$. Since the OABM parameters are defined on $(0,\infty)$, the former families apply directly. As we show below, the fitted mean lies more than three standard deviations above zero, so the probability of $x<0$ is below 0.1\% and the Gaussian remains a valid approximation. 

Each fitted distribution was evaluated using the Bayesian information criterion (BIC) \cite{Konishi2008info},
\begin{equation}
\mathrm{BIC}=k\ln N-2\ln \widehat{\mathcal{L}},
\end{equation}
where $k$ is the number of fitted parameters, $N$ is the number of data points, and $\widehat{\mathcal{L}}$ is the maximized likelihood. Lower BIC values indicate a better balance between goodness of fit and model complexity. Since the three candidate distributions considered here have the same number of fitted parameters, differences in BIC are determined mainly by differences in the maximized likelihood.  

For the swimming velocity $V$ in the control group, the log-normal distribution gave the lowest BIC value. We therefore selected the log-normal distribution for $V$ under control conditions. However, in the hyperactivated group, the normal distribution produced the lowest BIC value. Using maximum likelihood estimation (MLE), we obtained $\mu = 4.44$ when $V$ has units of $\mu$m/s) and $\sigma = 0.42$ for the control group, and $\chi = 179.3$~$\mu$m/s and $\nu = 56.7$~$\mu$m/s for the hyperactivated group. 

For the angular frequency $\Omega$, the Gamma distribution was selected for the control group. Whereas the log-normal distribution produced a lower BIC value for the hyperactivated group. Using MLE, we obtained $\beta = 0.07$~s and $\lambda = 4.0$ for the control group, and $\mu = 4.32$ and $\sigma = 0.44$ for the hyperactivated group.

For the oscillatory amplitude $A$, the Gaussian distribution has the lowest BIC. MLE yields $\mu = 0.51\pi$ and $\sigma = 0.12\pi$ for the control group, and $\mu = 0.53\pi$ and $\sigma = 0.13\pi$ for the hyperactivated group.

Finally, the rotational diffusion coefficient $D_R$ was estimated from the slow decay of the normalized VACF computed with the lag-dependent velocity $\mathbf v(t)$ at $\delta t=25\Delta t$. This value of $k$ suppresses most of the fast oscillatory contribution, so that the remaining VACF is dominated by the slow directional-decorrelation envelope. In our approximation, this envelope is given by $\widehat{C}_{\bf v}(\tau)\simeq 1-D_R\tau$. Thus, for each trajectory, $D_R$ was obtained from a linear fit of the normalized VACF envelope, with $D_R$ given by the negative slope. The fitting procedure was adaptive. First, the full normalized VACF was fitted with a straight line. If the fit had a negative slope and a coefficient of determination $R^2\geq0.95$, the full fit was used. Otherwise, the local maxima of the VACF were identified and fitted instead, providing an estimate of the upper envelope. At least three local maxima were required for this envelope fit. If fewer than three maxima were available, the full linear fit was used as a fallback. This procedure allowed us to estimate $D_R$ consistently across trajectories whose VACFs ranged from nearly monotonic to weakly oscillatory. Using this procedure, the rotational diffusion coefficient $D_R$ in both groups was found to follow an exponential distribution \cite{carlton2017proba},
\begin{equation}
p(D_R) = \frac{1}{\langle D_R \rangle} \exp \bigg[ - \frac{D_R}{\langle D_R \rangle} \bigg],
\end{equation}
where $\langle D_R \rangle$ is the mean rotational diffusion coefficient. For this analysis, we considered only positive values of the rotational diffusion coefficient, $D_R > 0$. We then calculated the complementary cumulative distribution function (CCDF), shown in figure~\ref{fig:distribution}d of the main text, which for an exponential distribution is
\begin{equation}
P\Big(\hat{D}_R > \mathcal{D}_R \Big) = \exp\bigg[ - \frac{\mathcal{D}_R}{\langle D_R \rangle} \bigg].
\end{equation}
Therefore, on a semi-logarithmic scale, the CCDF is expected to follow a straight line with slope $-1/\langle D_R \rangle$. The mean rotational diffusion coefficients were found to be $\langle D_R \rangle = 1.64$ s$^{-1}$ for the control group and $\langle D_R \rangle = 1.19$ s$^{-1}$ for hyperactivation.


\section{Correlation among the OABM parameters \label{app:Correlation}}

\begin{table}[ht]
\centering
\begin{tabular}{lcccc}
\hline
 & $V$ & $A$ & $\Omega$ & $D_R$ \\
\hline
$V$      & 1.00 & 0.32 & 0.09 & $0.02$ \\
$A$      &  & 1.00 & $-0.11$  & 0.26 \\
$\Omega$ &  &  & 1.00 & $-0.12$ \\
$D_R$    &  &  &  & 1.00 \\
\hline
\end{tabular}
\caption{Pearson correlation matrix for the fitted OABM parameters in the control group (n = 169).}
\label{tab:corr_control}
\end{table}

\begin{table}[ht]
\centering
\begin{tabular}{lcccc}
\hline
 & $V$ & $A$ & $\Omega$ & $D_R$ \\
\hline
$V$      & 1.00 & 0.60 & $-0.32$ & 0.26 \\
$A$      &  & 1.00 & $-0.34$ & 0.45 \\
$\Omega$ &  &  & 1.00 & $-0.16$ \\
$D_R$    &  &  &  & 1.00 \\
\hline
\end{tabular}
\caption{Pearson correlation matrix for the fitted OABM parameters in the hyperactivated group (n = 369).}
\label{tab:corr_8br}
\end{table}

\begin{figure*}[hbt!]
\centering
\includegraphics[width=\textwidth]{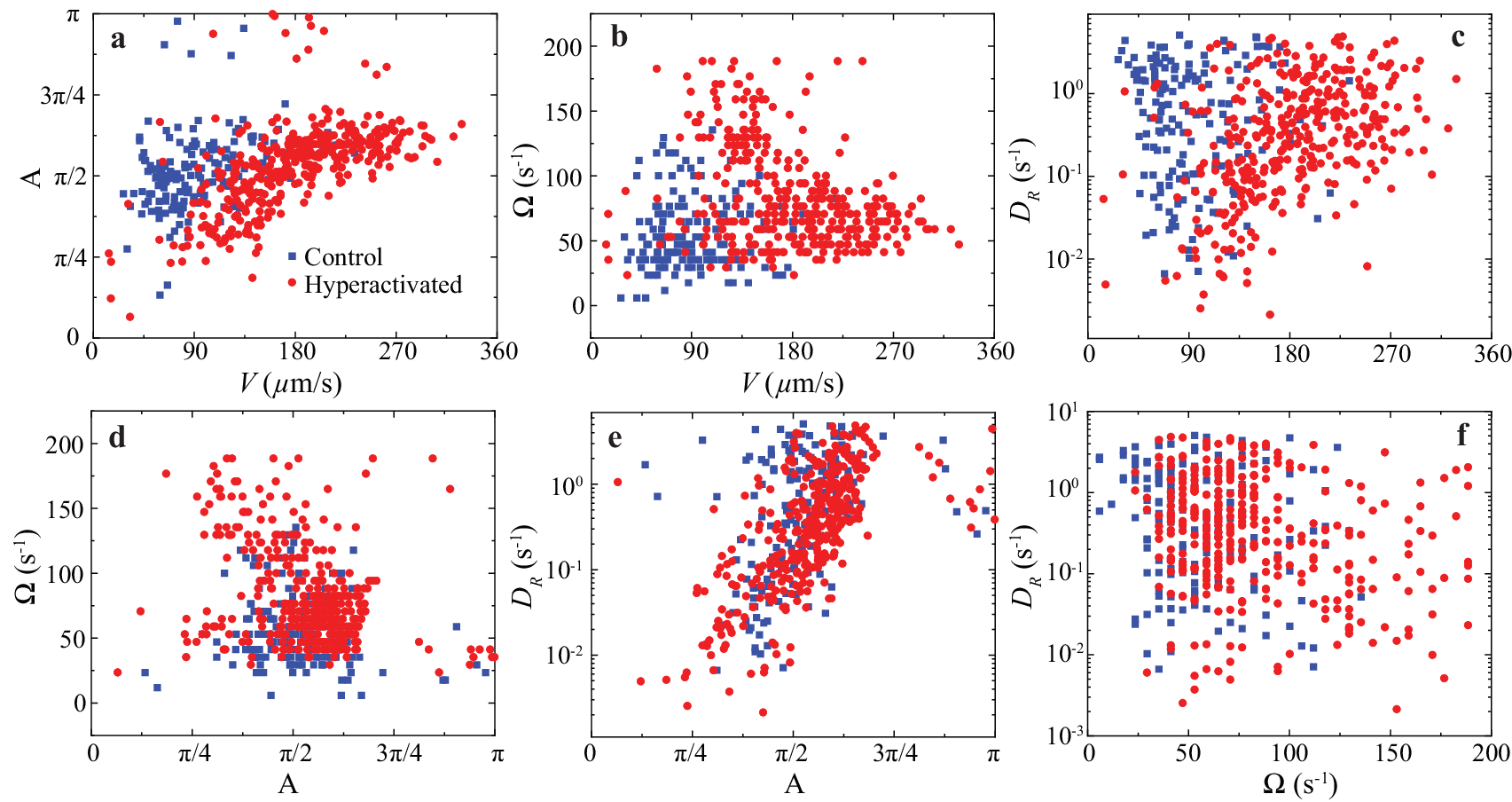}
\caption{\textbf{Pairwise correlations among the fitted OABM parameters for the control and hyperactivated groups.} Scatter plots are shown for (a) $A$ versus $V$, (b) $\Omega$ versus $V$, (c) $D_R$ versus $V$, (d) $\Omega$ versus $A$, (e) $D_R$ versus $A$, and (f) $D_R$ versus $\Omega$. The strongest and most relevant correlation is between $V$ and $A$.}
\label{fig:Correlations}
\end{figure*}

To assess whether the OABM parameters could be sampled independently at the population level, we computed the Pearson correlation coefficients between the fitted parameters $V$, $A$, $\Omega$, and $D_R$. The corresponding pairwise scatter plots are shown in figure~\ref{fig:Correlations}, and the correlation matrices are reported in Tables~\ref{tab:corr_control} and \ref{tab:corr_8br}. In both groups, the strongest correlation common to both experimental conditions is between $V$ and $A$. The corresponding coefficients are $\rho_{V,A}^{\rm C} = 0.32$ and $\rho_{V,A}^{\rm H}=0.60$. Therefore, the inferred values of $A$ are not statistically independent of the measured velocities $V$. Sampling $V$ and $A$ independently would remove this empirical dependence.
A smaller but significant positive correlation between $A$ and $D_R$ is also observed. However, this dependence was not included in the minimal population-level model because the trajectories are analyzed over time scales short compared with the slow orientational decorrelation time. In this regime, $D_R$ mainly affects the slow decorrelation envelope, whereas the leading short-time and oscillatory behavior is controlled by $V$, $A$, and $\Omega$.
The hyperactivated group also shows moderate negative correlations involving $\Omega$, particularly with $V$ and $A$. These correlations were not included in the minimal sampling scheme. We retained only the $V$-$A$ dependence because it is the strongest correlation common to both experimental conditions and is directly linked to the amplitude-estimation procedure.

\bibliographystyle{unsrt}
\bibliography{sample}

\end{document}